\documentclass[review,sort&compress,12pt]{elsarticle}

\usepackage{lineno,hyperref}
\usepackage{graphicx}
\usepackage{booktabs}
\usepackage{amssymb}
\usepackage{amsmath}
\usepackage{color,soul}
\usepackage{subfigure}

\begin{document}
	
	\begin{frontmatter}
		
		
		

		
		
		\title{Diffraction of a symmetric TM mode at an open-ended deeply corrugated waveguide with a small period}
		
		\author[spbu]{Evgenii S. Simakov\corref{mycorrespondingauthor}}
		\cortext[mycorrespondingauthor]{Corresponding author}
		\ead{st040497@student.spbu.ru}
		
		\author[spbu]{Sergey N. Galyamin}
		
		\author[spbu]{Andrey V. Tyukhtin}
		
		\address[spbu]{Saint Petersburg State University, 7/9 Universitetskaya nab., St. Petersburg, 199034 Russia}
		
		\begin{abstract}
			We investigate the diffraction of a slow symmetric TM mode by an open-ended corrugated cylindrical waveguide with a flange. 
			This mode can be generated, in particular, by a charged particle bunch moving along the waveguide axis. 
			We analyze the so-called ''longwave'' range when the wavelengths and the waveguide radius are much greater than the corrugation period. 
			In this case, the corrugated waveguide wall can be replaced with a smooth one on which the so-called equivalent boundary conditions (EBC) are fulfilled. 
			Here, we also assume that the wall is deeply corrugated, i.e. the corrugation depth is of the same order as an inverse wavenumber.
			The approach applied uses the solution of the corresponding Wiener–Hopf–Fock equation. 
			This solution is in turn utilized to construct an infinite linear system for reflection coefficients which can be solved numerically using the reduction technique. 
			The dependences of the components of the electromagnetic field on the problem parameters are obtained and analyzed.
			Typical radiation patterns are presented. 
		\end{abstract}
		
	\end{frontmatter}
	
	\begin{keyword}
		corrugated structure, charged particle bunch, radiation of particle bunch, equivalent boundary conditions, surface wave, bunch diagnostics 
	\end{keyword}
	
	
	\section{Introduction} 
	
	The interaction of an electromagnetic field from charged particles with periodic structures is commonly used as a source of Smith-Purcell radiation (SPR).
	This is the well-known physical phenomenon which is, by its nature, the diffraction radiation at the wavelengths comparable to a structure period~%
	\cite{Pot1,Pot2}.
	In contrast to SPR, electromagnetic radiation in the so-called ``longwave'' range is only gaining popularity in the scientific community.
	Under the term ``longwave'' we mean that the wavelengths considered are much greater than a structure period.
	The radiation at these wavelengths is connected with a specific anisotropy of a periodic structure and therefore has a principal difference from SPR.
	Studying the ``longwave'' radiation can be of great importance in accelerator and beam physics.
	In particular, this can be used for the generation of powerful THz emission~%
	\cite{BaneStup3,BaneStup4}.
	
	The present work focuses just on the “longwave” case.
	The analytical solution of such problems can be obtained by the use of the so-called equivalent boundary conditions (EBC)~%
	\cite{NefSiv}.
	According to this method, a corrugated structure is replaced with a smooth (anisotropic) surface characterizing by a certain matrix impedance.
	Recently, we applied the EBC method in a series of the ``longwave'' problems where the structures with rectangular corrugation were analyzed.
	One can mention papers~%
	\cite{AVAkh,GrigCorSmooth,GrigOpEndArx,ESAV1,ESAV2} in which we study the shallow corrugated structures.
	The term ``shallow'' means that the corrugation depth is much less than the wavelengths under consideration.
	In particular, paper~%
	\cite{GrigCorSmooth} is devoted to studying the wave field from a charged particle bunch intersecting a boundary between the corrugated waveguide area and an area with a smooth wall.
	It is worth noting that this work contains the results of numerical simulations in CST Particle Studio which demonstrate the correctness of the EBC method. 
	As well, we should note papers~%
	\cite{ESAV3,ESAV4} where the deeply corrugated planar structure is considered.
	The deep corrugation, unlike the shallow one, is characterized by the corrugation depth which is of the same order as an inverse wavenumber.
	In~%
	\cite{ESAV3,ESAV4}, we examine two cases of a bunch motion: along the structure and perpendicularly to it.
	In both cases, the emphasis is on the effect of generating surface waves.
	In particular, we analyze the configuration of these waves and its possible application in bunch diagnostics.
	
	In the present paper, we investigate the diffraction of a slow symmetric TM mode at an open-end of a corrugated cylindrical waveguide with a flange.
	The waveguide wall is assumed to be perfectly conductive and has the rectangular corrugation.
	The period of the corrugation is much less than the waveguide radius and the wavelengths under consideration (the ``longwave'' case).
	The depth of the corrugation is of the same order as an inverse wavenumber (the deep corrugation).
	The incident TM mode is excited by a charged particle bunch moving along the waveguide axis.
	The electromagnetic field of the incident mode is obtained by using the EBC method.
	The reflected field inside the waveguide is decomposed into a series of waveguide eigenmodes.
	The field outside the waveguide (in free space) is presented by integral transforms. 
	The solution of the corresponding Wiener-Hopf-Fock equation is obtained using the factorization method~%
	\cite{WHFWeinstein,MittraLee} and then utilized to construct an infinite linear system for the reflection coefficients of the waveguide modes.
	
	It should be noted that the method based on Wiener-Hopf-Fock technique and presented in this work was previously used in a series of problems concerning semi-infinite open-ended waveguides.
	For example, paper~%
	\cite{VoskZh1} contains the solution for a planar waveguide with a flange (the waveguide wall is smooth).
	In~%
	\cite{IEEEopendw1,GalOpEndArx}, the approach is generalized to a cylindrical waveguide with uniform and layered dielectric filling.
	In~%
	\cite{GrigOpEndArx}, the solution is constructed for the cylindrical waveguide having the shallow corrugated wall (the vacuum case and uniform dielectric filling are considered).
	As well, we should emphasize that paper~%
	\cite{IEEEopendw1} includes the comparison between the rigorous analytical approach and simulations carried out in COMSOL Multiphysics.
	The agreement can be considered excellent, which proves the theory presented.

	\section{Field inside the waveguide}

	\begin{figure}[h]
		\centering
		\includegraphics[]{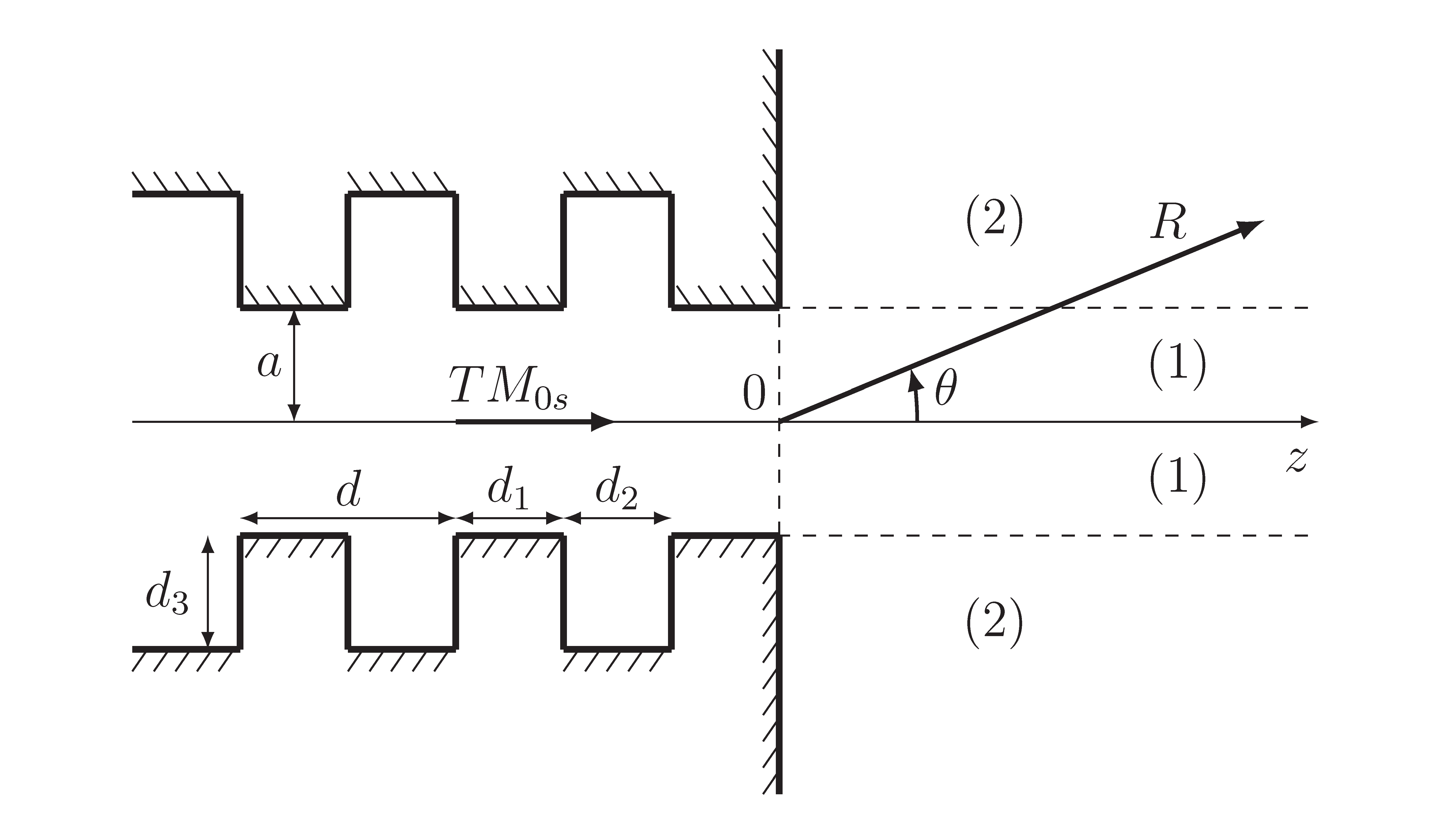}
		\caption[]{The open-ended deeply corrugated cylindrical waveguide with a flange.}
		\label{Fig:1} 
	\end{figure}
	A harmonic transverse magnetic mode TM$_{0s}$ falls on an open end of a cylindrical corrugated waveguide (Fig.~%
	\ref{Fig:1}).
	We assume that in the plane $z=0$, when $r>a$, the waveguide has an infinite flange. 
	The waveguide wall is perfectly conductive and has the rectangular corrugation.
	The period of the corrugation $d$ is much less than wavelength $\lambda$ and waveguide radius $a$:
	\begin{equation}
		\label{eq:2.1}
		\tag{2.1}
		d\ll\lambda,\;\;d\ll a.
	\end{equation}
	In this approximation, the corrugated wall can be replaced with a smooth surface on which the so-called equivalent boundary conditions (EBC) are fulfilled~%
	\cite{NefSiv}.
	
	Here, we also assume that the corrugation is deep, i.e.
	\begin{equation}
		\label{eq:2.2}
		\tag{2.2}
		k_0d_3\sim 1,
	\end{equation}
	where $d_3$ is the corrugation depth, $k_0=\omega/c=2\pi/\lambda$ is a wavenumber.
	Under condition~%
	\eqref{eq:2.2}, the EBC for the Fourier transforms of electric and magnetic fields are the following~%
	\cite{NefSiv}:
	\begin{equation}
		\label{eq:2.3}
		\tag{2.3}
		E_{z\omega}\big|_{r=a}=i\eta H_{\varphi\omega}\big|_{r=a}, \; E_{\varphi\omega}\big|_{r=a}=0,
	\end{equation}
	where impedance $\eta$ has the form
	\begin{equation}
		\label{eq:2.4}
		\tag{2.4}
		\eta=\frac{d_2}{d}\frac{\operatorname{tg}\left(k_0d_3\right)}{1-k_0d\;l\operatorname{tg}\left(k_0d_3\right)}.
	\end{equation}
	Parameter $l$ in~%
	\eqref{eq:2.4} is
	positive and quite small: $0<l\lesssim0.082$ (see~%
	\cite{ESAV3,NefSiv} for details).
	
	The incident TM$_{0s}$ mode can be generated, in particular, by a charged particle bunch moving along the waveguide axis and characterized by the charge density $\rho=q\delta\left(\mathbf{r}\right)\kappa\left(\zeta\right)$, where $\zeta=z-Vt$, $\kappa\left(\zeta\right)$ is some longitudinal charge distribution, $\mathbf{V}=c\beta\mathbf{e}_z$ is a bunch velocity.  
	The components of this mode are
	\begin{equation}
		\label{eq:2.5}
		\tag{2.5}
		H_{\varphi}^{(i)}=-\frac{iqk_{0s}\tilde\kappa}{a}M^{(i)}I_1\left(\tilde\chi_sr\right)\exp\left(ik_{0s}\beta^{-1}z\right), 
	\end{equation}
    \begin{equation}
		\label{eq:2.6}
		\tag{2.6}
		E_{r}^{(i)}=-\frac{iqk_{0s}\tilde\kappa}{a\beta}M^{(i)}I_1\left(\tilde\chi_sr\right)\exp\left(ik_{0s}\beta^{-1}z\right), 
	\end{equation}
    \begin{equation}
		\label{eq:2.7}
		\tag{2.7}
		E_{z}^{(i)}=\frac{qk_{0s}\tilde\kappa}{a\beta\gamma}M^{(i)}I_0\left(\tilde\chi_sr\right)\exp\left(ik_{0s}\beta^{-1}z\right).
	\end{equation}
	We note that, for the sake of convenience, the electromagnetic field of the incident mode is written in a complex form.
	A real (physically measurable) field can be found as the real part of~%
	\eqref{eq:2.5}-%
	\eqref{eq:2.7}.
	In these formulas, 
	\begin{equation}
		\label{eq:2.8}
		\tag{2.8}
		\tilde\kappa=\frac{1}{2\pi}\int\limits_{-\infty}^{+\infty}d\zeta\kappa\left(\zeta\right)\exp\left(-ik_{0s}\beta^{-1}\zeta\right)
	\end{equation}
	is the Fourier-transform of longitudinal charge distribution $\kappa\left(\zeta\right)$, coefficient $M^{(i)}$ has the form
	\begin{align}
		\label{eq:2.9} 
		\tag{2.9}
		M^{(i)}{=}{-}\frac{8\pi\left[K_0\left(\tilde\chi_sa\right)+\beta\gamma\eta K_1\left(\tilde\chi_sa\right)\right]}{\beta\gamma\eta I_0\left(\tilde\chi_sa\right){-}I_1\left(\tilde\chi_sa\right)\left[1{+}\beta\gamma \left(a\tilde\chi_s\right)^{-1}\left(\eta-k_{0s}\frac{\partial\eta}{\partial k_0}\big|_{k_0=k_{0s}}\right)\right]},
	\end{align}
	where $\gamma=\left(1-\beta^2\right)^{-1/2}$, $K_{0,1}\left(\tilde\chi_sa\right)$ are Macdonald functions, $I_{0,1}\left(\tilde\chi_sr\right)$ are modified Bessel functions, $\tilde\chi_s=|\chi_s|=k_{0s}\left(\beta\gamma\right)^{-1}$, $k_{0s}=\omega_s /c$ is a wavenumber which is found from the dispersion equation
	\begin{equation}
		\label{eq:2.10}
		\tag{2.10}
		I_0\left(\frac{k_0a}{\beta\gamma}\right)-\beta\gamma\eta I_1\left(\frac{k_0a}{\beta\gamma}\right) = 0.
	\end{equation}

    The reflected field inside the waveguide is presented as a series of eigenmodes propagating in the opposite direction:
	\begin{equation}
		\label{eq:2.11}
		\tag{2.11}
		H_{\varphi}^{(r)}=-\frac{iqk_{0s}}{a}\sum_{m=1}^{\infty}M_mJ_1\left(\chi_mr\right)\exp\left(-ik_{zm}z\right), 
	\end{equation}
    \begin{equation}
		\label{eq:2.12}
		\tag{2.12}
		E_{r}^{(r)}=\frac{iq}{a}\sum_{m=1}^{\infty}k_{zm}M_mJ_1\left(\chi_mr\right)\exp\left(-ik_{zm}z\right), 
	\end{equation}
    \begin{equation}
		\label{eq:2.13}
		\tag{2.13}
		E_{z}^{(r)}=\frac{q}{a}\sum_{m=1}^{\infty}\chi_mM_mJ_0\left(\chi_mr\right)\exp\left(-ik_{zm}z\right),
	\end{equation}
	where $J_{0,1}$ are Bessel functions and $\left\{M_m\right\}$ are unknown coefficients which will be determined later.
	Transverse wavenumbers $\left\{\chi_m\right\}$ are found from equivalent boundary conditions~%
	\eqref{eq:2.3}.
	The relationship between $\left\{\chi_m\right\}$ and longitudinal wavenumbers $\left\{k_{zm}\right\}$ is given by  $k_{zm}=\sqrt{k_{0s}^2 - \chi_m^2}$, $\operatorname{Im}k_{zm}>0$.

	\section{Field outside the waveguide}
	
	Below, we will consider only the component of the magnetic field $H_{\varphi}$.
	Expressions for the components of the electric field $E_{r}$ and $E_{z}$ are easy to obtain by the formulas
	\begin{align}
		\label{eq:3.1}
		\tag{3.1}
		E_{r}=\frac{1}{ik_{0s}}\frac{\partial H_{\varphi}}{\partial z},\;\;\;E_{z}=\frac{-1}{ik_{0s}r}\left[\frac{\partial}{\partial r}\left(rH_{\varphi}\right)\right].
	\end{align}

	The area outside the waveguide ($z>0$) is divided into two parts: $r<a$ and $r>a$ (they are designated by $(1)$ and $(2)$ in Fig.~%
	\ref{Fig:1}). 
	The magnetic field in areas $(1)$ and $(2)$ is described by the Helmholtz equation
	\begin{equation}
		\label{eq:3.2}
		\tag{3.2}
		\left[\frac{\partial^2}{\partial z^2}+\frac{1}{r}\frac{\partial}{\partial r}\left(r\frac{\partial}{\partial r}\right)-\frac{1}{r^2}+k_{0s}^2\right]H_{\varphi}^{(1,2)}=0.
	\end{equation}
	For each area, we introduce functions $\Psi_{\pm}^{(2)}\left(r,\alpha\right)$ and $\Psi_+^{(1)}\left(r,\alpha\right)$ which are the following integral transformations:
	\begin{equation}
		\label{eq:3.3}
		\tag{3.3}
		\Psi_+^{(2)}\left(r,\alpha\right)=\frac{1}{2\pi}\int\limits_{0}^{+\infty}H_{\varphi}^{(2)}\left(r,z\right)\exp\left(i\alpha z\right)dz, 
    \end{equation}
    \begin{equation}
		\label{eq:3.4}
		\tag{3.4}
		\Psi_-^{(2)}\left(r,\alpha\right)=\frac{1}{2\pi}\int\limits_{-\infty}^{0}H_{\varphi}^{(2)}\left(r,-z\right)\exp\left(i\alpha z\right)dz,
	\end{equation}
    \begin{equation}
		\label{eq:3.5}
		\tag{3.5}
		\Psi_{+}^{(1)}\left(r,\alpha\right)=\frac{1}{2\pi}\int\limits_{0}^{+\infty}H_{\varphi}^{(1)}\left(r,z\right)\exp\left(i\alpha z\right)dz.
	\end{equation}
	In formulas~%
	\eqref{eq:3.3}-%
	\eqref{eq:3.5} and hereafter subscript ``$+$'' (or ``$-$'') means that the function is regular in the area $\operatorname{Im}\alpha>-\delta$ (or $\operatorname{Im}\alpha<\delta$), where $\delta>0$.
	Note that function $\Psi_+^{(2)}\left(r,-\alpha\right)=\Psi_-^{(2)}\left(r,\alpha\right)$ and therefore is regular in the area $\operatorname{Im}\alpha<\delta$.
	
	Applying integral operator $\left(2\pi\right)^{-1}\!\!\int_0^{+\infty}\!\cdot\exp\left(i\alpha z\right)dz$ to equation~%
	\eqref{eq:3.2} we have 
	\begin{align}
		\notag
		\left[\frac{1}{r}\frac{\partial}{\partial r}\left(r\frac{\partial}{\partial r}\right)-\frac{1}{r^2}+k_{0s}^2\right]\Psi_+^{(1,2)}\left(r,\alpha\right) \\
		\label{eq:3.6}
		\tag{3.6}
		+\frac{1}{2\pi}\int\limits_0^{+\infty}\left(\frac{\partial^2}{\partial z^2}H_{\varphi}^{(1,2)}\right)&\exp\left(i\alpha z\right)dz=0.
	\end{align}
	Integrating the second term in~%
	\eqref{eq:3.6} by parts twice we obtain
	\begin{align}
		\notag
		&\left[\frac{1}{r}\frac{\partial}{\partial r}\left(r\frac{\partial}{\partial r}\right)-\frac{1}{r^2}+\varkappa^2\right]\Psi_+^{(1,2)}\left(r,\alpha\right) \\
		\label{eq:3.7}
		\tag{3.7}
		&=-\frac{1}{2\pi}\left[\exp\left(i\alpha z\right)\frac{\partial H_{\varphi}^{(1,2)}}{\partial z}\bigg|_{0}^{+\infty}-i\alpha\exp\left(i\alpha z\right)H_{\varphi}^{(1,2)}\bigg|_{0}^{+\infty}\right],
	\end{align}
	where
	\begin{equation}
		\label{eq:3.8}
		\tag{3.8}
		\varkappa=\sqrt{k_{0s}^2-\alpha^2},\;\;\;\operatorname{Im}\varkappa>0.
	\end{equation}

	\subsection{Equation for function $\Psi^{(2)}\left(r,\alpha\right)$}
	
	First, we consider area $(2)$.
	Taking into account that the waveguide flange is perfectly conductive ($E_{r}^{(2)}=0$) we can write boundary conditions at $z=+0$ and $z=+\infty$ ($r>a$) as follows:
	\begin{equation}
		\label{eq:3.9}
		\tag{3.9}
		\frac{\partial H_{\varphi}^{(2)}}{\partial  z}\bigg|_{z=+0}=H_{\varphi}^{(2)}\bigg|_{z=+\infty}=0.
	\end{equation}
	Therefore,~%
	\eqref{eq:3.7} takes the form
	\begin{equation}
		\label{eq:3.10}
		\tag{3.10}
		\left[\frac{1}{r}\frac{\partial}{\partial r}\left(r\frac{\partial}{\partial r}\right)-\frac{1}{r^2}+\varkappa^2\right]\Psi_+^{(2)}\left(r,\alpha\right)=-\frac{i\alpha}{2\pi}H_{\varphi}^{(2)}\bigg|_{z=+0},
	\end{equation}
	The sum of~%
	\eqref{eq:3.10} and the same equation for $-\alpha$ results in
	\begin{equation}
		\label{eq:3.11}
		\tag{3.11}
		\left[\frac{1}{r}\frac{\partial}{\partial r}\left(r\frac{\partial}{\partial r}\right)-\frac{1}{r^2}+\varkappa^2\right]\left[\Psi^{(2)}_+\left(r,\alpha\right)+\Psi_-^{(2)}\left(r,\alpha\right)\right]=0.
	\end{equation}
	This is Bessel equation which has the solution 
	\begin{equation}
		\label{eq:3.12}
		\tag{3.12}
		\Psi^{(2)}\left(r,\alpha\right)\equiv \Psi_+^{(2)}\left(r,\alpha\right)+\Psi_-^{(2)}\left(r,\alpha\right)=C_2H_1^{(1)}\left(\varkappa r\right),
	\end{equation}
	where $H_1^{(1)}\left(\varkappa r\right)$ is Hankel function of the first kind, $C_2$ is a coefficient which will be defined later.

	\subsection{Equation for function $\Psi_{+}^{(1)}\left(r,\alpha\right)$}
	
	Now, we consider area $(1)$.
	Boundary conditions for the magnetic field in this area can be written in the form
	\begin{equation}
		\label{eq:3.13}
		\tag{3.13}
		H_{\varphi}^{(1)}\bigg|_{z=+\infty}=\frac{\partial H_{\varphi}^{(1)}}{\partial z}\bigg|_{z=+\infty}=0, 
	\end{equation}
    \begin{equation}
		\label{eq:3.14}
		\tag{3.14}
		\left(H_{\varphi}^{(i)}+H_{\varphi}^{(r)}\right)\bigg|_{z=-0}=H_{\varphi}^{(1)}\bigg|_{z=+0}, 
	\end{equation}
    \begin{equation}
		\label{eq:3.15}
		\tag{3.15}
		\left(\frac{\partial H_{\varphi}^{(i)}}{\partial  z}+\frac{\partial H_{\varphi}^{(r)}}{\partial z}\right)\bigg|_{z=-0}=\frac{\partial H_{\varphi}^{(1)}}{\partial z}\bigg|_{z=+0}.
	\end{equation}
	Conditions~%
	\eqref{eq:3.14} and~%
	\eqref{eq:3.15} mean the continuity of components $H_{\varphi}$ and $E_{r}$, correspondingly, in the plane $z=0$ ($r<a$).
	Taking into consideration~%
	\eqref{eq:3.13}-%
	\eqref{eq:3.15} and formulas for the magnetic field inside the waveguide~%
	\eqref{eq:2.5} and~%
	\eqref{eq:2.11} we derive from~%
	\eqref{eq:3.7} the expression
	\begin{align}
		\label{eq:3.16}
		&\left[\frac{1}{r}\frac{\partial}{\partial r}\left(r\frac{\partial}{\partial r}\right)-\frac{1}{r^2}+\varkappa^2\right]\Psi_+^{(1)}\left(r,\alpha\right)= \nonumber \\
		\tag{3.16}
		&\frac{qk_{0s}}{2\pi a}\left[\left(\frac{k_{0s}}{\beta}-\alpha\right)\tilde\kappa M^{(i)}I_1\left(\tilde\chi_sr\right)-\sum_{m=1}^{\infty}\left(k_{zm}+\alpha\right)M_mJ_1\left(\chi_mr\right)\right].
	\end{align}
	A solution of~%
	\eqref{eq:3.16} is a sum of a general solution of a homogeneous equation $\Psi_g^{(1)}\left(r,\alpha\right)$ and a particular solution of an inhomogeneous equation $\Psi_p^{(1)}\left(r,\alpha\right)$. 
	The homogeneous equation is the first-order Bessel equation, and therefore one can write the general solution in the form
	\begin{align}
		\label{eq:3.17}
		\tag{3.17}
		\Psi_g^{(1)}\left(r,\alpha\right)=C_1J_1\left(\varkappa r\right).
	\end{align}
	According to the right part of~%
	\eqref{eq:3.16}, the particular solution can be presented as a series of Bessel functions
	\begin{align}
		\label{eq:3.18}
		\tag{3.18}
		\Psi_{p}^{(1)}\left(r,\alpha\right)=AI_1\left(\tilde\chi_sr\right)+\sum_{m=1}^{\infty}B_mJ_1\left(\chi_mr\right).
	\end{align}
	Substituting~%
	\eqref{eq:3.18} in~%
	\eqref{eq:3.16} and performing some mathematical transformations we obtain the following expressions for the coefficients $A, \left\{B_m\right\}$:
	\begin{equation}
		\label{eq:3.19}
		\tag{3.19}
		A=\frac{qk_{0s}}{2\pi a}\frac{\tilde\kappa M^{(i)}}{k_{0s}\beta^{-1}+\alpha},\;\;\;B_m=-\frac{qk_{0s}}{2\pi a}\frac{M_m}{k_{zm}-\alpha}.
	\end{equation}
	Thus, the solution of~%
	\eqref{eq:3.16} is
	\begin{align}
		\label{eq:3.20}
		\tag{3.20}
		\Psi_+^{(1)}\left(r,\alpha\right)=C_1J_1\left(\varkappa r\right)+\frac{qk_{0s}}{2\pi a}\left[\frac{\tilde\kappa M^{(i)}I_1\left(\tilde\chi_sr\right)}{k_{0s}\beta^{-1}+\alpha}-\sum_{m=1}^{\infty}\frac{M_mJ_1\left(\chi_mr\right)}{k_{zm}-\alpha}\right].
	\end{align}

	\subsection{Functions $\Phi_{+}^{(1)}\left(r,\alpha\right)$, $\Phi_{\pm}^{(2)}\left(r,\alpha\right)$}
	
	Let us introduce new functions
	\begin{equation}
		\label{eq:3.21}
		\tag{3.21}
		\Phi_+^{(2)}\left(r,\alpha\right)=\frac{1}{2\pi}\int\limits_{0}^{+\infty}E_{z}^{(2)}\left(r,z\right)\exp\left(i\alpha z\right)dz, 
	\end{equation}
    \begin{equation}
		\label{eq:3.22}
		\tag{3.22}
		\Phi_-^{(2)}\left(r,\alpha\right)=\frac{1}{2\pi}\int\limits_{-\infty}^{0}E_{z}^{(2)}\left(r,-z\right)\exp\left(i\alpha z\right)dz,
	\end{equation}
    \begin{equation}
		\label{eq:3.23}
		\tag{3.23}
		\Phi_{+}^{(1)}\left(r,\alpha\right)=\frac{1}{2\pi}\int\limits_{0}^{+\infty}E_{z}^{(1)}\left(r,z\right)\exp\left(i\alpha z\right)dz.
	\end{equation}
	As stated above, subscripts ``$+$'' and ``$-$'' mean that the function is regular in the area $\operatorname{Im}\alpha>-\delta$ and $\operatorname{Im}\alpha<\delta$, correspondingly.
	Taking into consideration the second expression in~%
	\eqref{eq:3.1} and formulas ~%
	\eqref{eq:3.12},~%
	\eqref{eq:3.20} we obtain for area $(1)$
	\begin{align}
		\label{eq:3.24}
		\tag{3.24}
		\Phi_{+}^{(1)}\!\left(r,\alpha\right){=}\frac{i\varkappa C_1}{k_{0s}}J_0\!\left(\varkappa r\right){+}\frac{iq}{2\pi a}\!\!\left[\!\frac{\tilde\chi_s\tilde\kappa M^{(i)}}{k_{0s}\beta^{-1}{+}\alpha}I_0\!\left(\tilde\chi_sr\right){-}\!\!\sum_{m=1}^{\infty}\!\frac{\chi_m M_m}{k_{zm}{-}\alpha}J_0\!\left(\chi_mr\right)\right]\!
	\end{align}
	and for area $(2)$
	\begin{equation}
		\label{eq:3.25}
		\tag{3.25}
		\Phi_{+}^{(2)}\left(r,\alpha\right)+\Phi_{-}^{(2)}\left(r,\alpha\right)=\frac{i\varkappa}{k_{0s}}C_2H_0^{(1)}\left(\varkappa r\right).
	\end{equation}

	\section{Wiener-Hopf-Fock equation}
	
	We note that the continuity of $E_{z}$ and $H_{\varphi}$ at $r=a$ ($z>0$) is expressed through conditions
	\begin{equation}
		\label{eq:4.1}
		\tag{4.1}
		\Phi_+^{(1)}\left(a,\alpha\right)=\Phi_+^{(2)}\left(a,\alpha\right),
	\end{equation}
    \begin{equation}
		\label{eq:4.2}
		\tag{4.2}
		\Psi_+^{(1)}\left(a,\alpha\right)=\Psi_+^{(2)}\left(a,\alpha\right).
	\end{equation}
	Coefficient $C_1$ can be found from~%
	\eqref{eq:4.1} as follows:
	\begin{align}
		\label{eq:4.3}
		&C_1=\frac{k_{0s}}{i\varkappa J_0\left(\varkappa a\right)}\Phi_+^{(2)}\left(a,\alpha\right)- \nonumber \\
		\tag{4.3}
		&\frac{qk_{0s}}{2\pi a\varkappa J_0\left(\varkappa a\right)}\left[\frac{\tilde\chi_s\tilde\kappa M^{(i)}}{k_{0s}\beta^{-1}+\alpha}I_0\left(\tilde\chi_sa\right)-\sum_{m=1}^{\infty}\frac{\chi_m M_m}{k_{zm}-\alpha}J_0\left(\chi_ma\right)\right].
	\end{align}
	The substitution of~%
	\eqref{eq:4.3} in~%
	\eqref{eq:4.2} results in the expression
	\begin{align}
		\label{eq:4.4}
		&\Psi_+^{(2)}\left(a,\alpha\right)=-\frac{k_{0s}}{J_0\left(\varkappa a\right)} \nonumber \\
		\tag{4.4}
		&\left\{\frac{i}{\varkappa}\Phi_+^{(2)}\left(a,\alpha\right)J_1\left(\varkappa a\right)+\frac{q}{2\pi a}\left[\frac{\tilde\kappa M^{(i)}\xi_s\left(a,\alpha\right)}{k_{0s}\beta^{-1}+\alpha}-\sum_{m=1}^{\infty}\frac{M_m\xi_m\left(a,\alpha\right)}{k_{zm}-\alpha}\right]\right\},
	\end{align}
	where
	\begin{equation}
		\label{eq:4.5}
		\tag{4.5}
		\xi_s\left(a,\alpha\right)=\frac{\tilde\chi_s}{\varkappa}J_1\left(\varkappa a\right)I_0\left(\tilde\chi_s a\right)-I_1\left(\tilde\chi_s a\right)J_0\left(\varkappa a\right), 
	\end{equation}
    \begin{equation}
		\label{eq:4.6}
		\tag{4.6}
		\xi_m\left(a,\alpha\right)=\frac{\chi_m}{\varkappa}J_1\left(\varkappa a\right)J_0\left(\chi_m a\right)-J_1\left(\chi_m a\right)J_0\left(\varkappa a\right).
	\end{equation}
	Note that~%
	\eqref{eq:4.4} has singularities when $J_0\left(j_{0n}\right)=0$, where $j_{0n}=\varkappa_n a$ is the $n$th zero of Bessel function $J_0$.
	According to~%
	\eqref{eq:3.8}, $\alpha_n$ is given by
	\begin{equation}
		\label{eq:4.7}
		\tag{4.7}
		\alpha_n=\sqrt{k_{0s}^2-\left(\frac{j_{0n}}{a}\right)^2},\;\;\;n=1,2,\dots
	\end{equation}
	The function on the left-hand side of~%
	\eqref{eq:4.4} is free of poles in the upper half-plane of $\alpha$.
	To eliminate poles~%
	\eqref{eq:4.7} on the right-hand side of~%
	\eqref{eq:4.4} one should require $\left\{\dots\right\}=0$ at $\alpha=\alpha_n$:
	\begin{equation}
		\label{eq:4.8}
		\tag{4.8}
		\frac{i}{\varkappa_n}\Phi_+^{(2)}\left(a,\alpha_n\right)J_1\left(\varkappa_n a\right){+}\frac{q}{2\pi a}\left[\frac{\tilde\kappa M^{(i)}\xi_s\left(a,\alpha_n\right)}{k_{0s}\beta^{-1}+\alpha_n}{-}\!\!\sum_{m=1}^{\infty}\frac{M_m\xi_m\left(a,\alpha_n\right)}{k_{zm}-\alpha_n}\right]\!\!{=}0.
	\end{equation}

	Then, we express coefficient $C_2$ from~%
	\eqref{eq:3.25} and substitute it in~%
	\eqref{eq:3.12} assuming that $r=a$:
	\begin{equation}
		\label{eq:4.9}
		\tag{4.9}
		\Psi_+^{(2)}\left(a,\alpha\right)+\Psi_-^{(2)}\left(a,\alpha\right)=-\frac{ik_{0s}H_1^{(1)}\left(\varkappa a\right)}{\varkappa H_0^{(1)}\left(\varkappa a\right)}\left[\Phi_+^{(2)}\left(a,\alpha\right)+\Phi_-^{(2)}\left(a,\alpha\right)\right].
	\end{equation}
    Using equation~%
    \eqref{eq:4.4} and the properties $\Psi_+\left(-\alpha\right){=}\Psi_-\left(\alpha\right)$, $\Phi_+\left(-\alpha\right){=}\Phi_-\left(\alpha\right)$ we obtain from~%
    \eqref{eq:4.9}
	\begin{equation}
		\label{eq:4.10}
		\tag{4.10}
		\Phi_+^{(2)}\left(a,\alpha\right)+\Phi_-^{(2)}\left(a,\alpha\right)=\frac{q}{2}\left[\tilde\kappa M^{(i)}G_s\left(\alpha\right)-\sum_{m=1}^{\infty}M_mG_m\left(\alpha\right)\right].
	\end{equation}
	This is Wiener-Hopf-Fock equation of the problem.
	Functions $G_{s,m}\left(\alpha\right)$ in~%
	\eqref{eq:4.10} are defined by the expressions
	\begin{equation}
		\label{eq:4.11}
		\tag{4.11}
		G_{s,m}\left(\alpha\right)=\varkappa^2H_0^{(1)}\left(\varkappa a\right)\xi_{s,m}\left(a,\alpha\right)F_{s,m}\left(\alpha\right),
	\end{equation}
	where
	\begin{equation}
		\label{eq:4.12}
		\tag{4.12}
		F_s\left(\alpha\right)=\frac{k_{0s}\beta^{-1}}{k_{0s}^2\beta^{-2}-\alpha^2},\;\;\;F_m\left(\alpha\right)=\frac{k_{zm}}{k_{zm}^2-\alpha^2}.
	\end{equation}
	Next, we decompose functions~%
	\eqref{eq:4.11} into sums
	\begin{equation}
		\label{eq:4.13}
		\tag{4.13}
		G_{s,m}\left(\alpha\right)=G_{s,m+}\left(\alpha\right)+G_{s,m-}\left(\alpha\right).
	\end{equation}
	where $G_{s,m+}\left(\alpha\right)$ and $G_{s,m-}\left(\alpha\right)$ are regular in the areas $\operatorname{Im}\alpha>-\delta$ and $\operatorname{Im}\alpha<\delta$, correspondingly (the details of such decomposition is given, for example, in~%
	\cite{MittraLee}).
	Substituting~%
	\eqref{eq:4.13} into~%
	\eqref{eq:4.10} we put ``$+$'' terms to the left-hand side of~%
	\eqref{eq:4.10} and ``$-$'' terms to the right-hand side. 
	This results in the equation
	\begin{align}
		\label{eq:4.14}
		&\Phi_+^{(2)}\left(a,\alpha\right)-\frac{q}{2}\left[\tilde\kappa M^{(i)}G_{s+}\left(\alpha\right)-\sum_{m=1}^{\infty}M_mG_{m+}\left(\alpha\right)\right] = \nonumber \\
		\tag{4.14}
		&-\Phi_-^{(2)}\left(a,\alpha\right)+\frac{q}{2}\left[\tilde\kappa M^{(i)}G_{s-}\left(\alpha\right)-\sum_{m=1}^{\infty}M_mG_{m-}\left(\alpha\right)\right]=P\left(\alpha\right),
	\end{align}
	where $P\left(\alpha\right)$ is a function regular in the whole complex plane of $\alpha$~%
	\cite{MittraLee}.
	Asymptotic estimations of all terms in~%
	\eqref{eq:4.14} show that $P\left(\alpha\right)\sim\alpha^{-1/2}$ when $|\alpha|\rightarrow\infty$.
	Therefore, $P\left(\alpha\right)=0$ due to Liouvilles's theorem.
	Thus, the solution of Wiener-Hopf-Fock equation is written as follows:
	\begin{equation}
		\label{eq:4.15}
		\tag{4.15}
		\Phi_+^{(2)}\left(a,\alpha\right)=\frac{q}{2}\left[\tilde\kappa M^{(i)}G_{s+}\left(\alpha\right)-\sum_{m=1}^{\infty}M_mG_{m+}\left(\alpha\right)\right].
	\end{equation}
	Functions $G_{s,m+}\left(\alpha\right)$ in~%
	\eqref{eq:4.15} are defined by the formulas~%
	\cite{MittraLee}
	\begin{equation}
		\label{eq:4.16}
		\tag{4.16}
		G_{s,m+}\left(\alpha\right)=\int\limits_{-\infty+ic}^{+\infty+ic}\frac{G_{s,m}\left(x\right)}{x-\alpha}dx,\;\;\;c\in\left(-\delta,\delta\right).
	\end{equation}

	Finally, we should determine unknown coefficients $\left\{M_m\right\}$.
	To do this, we substitute~%
	\eqref{eq:4.15} into~%
	\eqref{eq:4.8} and obtain the infinite linear system
	\begin{equation}
		\label{eq:4.17}
		\tag{4.17}
		\sum_{m=1}^{\infty}W_{nm}M_m=w_n,
	\end{equation}
	where
	\begin{equation}
		\label{eq:4.18}
		\tag{4.18}
		W_{nm}=\frac{a}{2\varkappa_n}J_1\left(\varkappa_n a\right)G_{m+}\left(\alpha_n\right)+\frac{\xi_m\left(a,\alpha_n\right)}{k_{zm}-\alpha_n},
    \end{equation}
    \begin{equation}
		\label{eq:4.19}
		\tag{4.19}
		w_n=\tilde\kappa M^{(i)}\left[\frac{a}{2\varkappa_n}J_1\left(\varkappa_n a\right)G_{s+}\left(\alpha_n\right)+\frac{\xi_s\left(a,\alpha_n\right)}{k_{0s}\beta^{-1}+\alpha_n}\right],
	\end{equation}
	and $n=1,2,\dots$
	System~%
	\eqref{eq:4.17} can be solved numerically by the use of the reducing technique.

	\section{Results and discussion}
	
	The Fourier-transform of the magnetic field in areas $(1)$ is defined by the inverse integral transformation
	\begin{equation}
		\label{eq:5.1}
		\tag{5.1}
		H_{\varphi}^{(1)}\left(r,z\right)=\int\limits_{-\infty}^{+\infty}\Psi_+^{(1)}\left(r,\alpha\right)\exp\left(-i\alpha z\right)d\alpha.
	\end{equation}
	Combining formulas~%
	\eqref{eq:3.20},~%
	\eqref{eq:4.3} and~%
	\eqref{eq:4.15} we obtain
	\begin{align}
		\notag
		\label{eq:5.2}
		&\Psi_+^{(1)}\left(r,\alpha\right){=}-\frac{qk_{0s}}{2\pi aJ_0\left(\varkappa a\right)}\left\{\frac{a}{2\varkappa}\left[\tilde\kappa M^{(i)}G_{s+}\left(\alpha\right){-}\sum_{m=1}^{\infty}M_mG_{m+}\left(\alpha\right)\right]J_1\left(\varkappa r\right){+}\right. \\ \nonumber 
		\tag{5.2}
		&\left.\frac{\tilde\kappa M^{(i)}\xi_s\left(r,\alpha\right)}{k_{0s}\beta^{-1}+\alpha}-\sum_{m=1}^{\infty}\frac{M_m\xi_m\left(r,\alpha\right)}{k_{zm}-\alpha}\right\},
	\end{align}
	where
	\begin{equation}
		\label{eq:5.3}
		\tag{5.3}
		\xi_s\left(r,\alpha\right)=\frac{\tilde\chi_s}{\varkappa}J_1\left(\varkappa r\right)I_0\left(\tilde\chi_s a\right)-I_1\left(\tilde\chi_s r\right)J_0\left(\varkappa a\right), \\
	\end{equation}
    \begin{equation}
		\label{eq:5.4}
		\tag{5.4}
		\xi_m\left(r,\alpha\right)=\frac{\chi_m}{\varkappa}J_1\left(\varkappa r\right)J_0\left(\chi_m a\right)-J_1\left(\chi_m r\right)J_0\left(\varkappa a\right).
	\end{equation}

	Concerning area $(2)$, we have the following integral transformation:
	\begin{equation}
		\label{eq:5.5}
		\tag{5.5}
		H_{\varphi}^{(2)}\left(r,z\right)=\int\limits_{-\infty}^{+\infty}\Psi^{(2)}\left(r,\alpha\right)\exp\left(-i\alpha z\right)d\alpha,
	\end{equation}
	where $\Psi^{(2)}\left(r,\alpha\right)$ is given by~%
	\eqref{eq:3.12}.
	Using~%
	\eqref{eq:3.25} we express $C_2$ and substitute it into~%
	\eqref{eq:3.12}.
	Taking into account formula~%
	\eqref{eq:4.10} we, finally, obtain
	\begin{align}
		\label{eq:5.6}
		&\Psi^{(2)}\left(r,\alpha\right)=-\frac{iqk_{0s}\varkappa}{2} \nonumber \\
		\tag{5.6}
		&\left[\tilde\kappa M^{(i)}\xi_{s}\left(a,\alpha\right)F_s\left(\alpha\right)-\sum_{m=1}^{\infty}M_m\xi_m\left(a,\alpha\right)F_m\left(\alpha\right)\right]H_1^{(1)}\left(\varkappa r\right).
	\end{align}
    Functions $\xi_{s,m}\left(a,\alpha\right)$, $F_{s,m}\left(\alpha\right)$ are given by~%
    \eqref{eq:4.5},~%
    \eqref{eq:4.6} and~%
    \eqref{eq:4.12}.
    We will investigate the behavior of integral~%
    \eqref{eq:5.5} in the far-field zone, i.e. when $k_{0s}R\gg1$, where $R=\sqrt{r^2+z^2}$ (see Fig.~%
    \ref{Fig:1}).
    In this situation, the saddle point method can be applied~%
    \cite{FelMarc,BleiHand}.
    We use standard substitutions
    \begin{equation}
    	\label{eq:5.7}
    	\tag{5.7}
    	r=R\operatorname{sin}\theta,\;\;\;z=R\operatorname{cos}\theta
    \end{equation}
    and introduce new variables
    \begin{equation}
    	\label{eq:5.8}
    	\tag{5.8}
    	\alpha=k_{0s}\operatorname{sin}\tau,\;\;\;\varkappa=k_{0s}\operatorname{cos}\tau.
    \end{equation}
    Performing asymptotic expansion of $H_1^{(1)}\left(\varkappa r\right)$ for $r|\varkappa|\gg1$ and using~%
    \eqref{eq:5.7},~%
    \eqref{eq:5.8} we obtain the exponential dependence as follows:
    \begin{equation}
    	\label{eq:5.9}
    	\tag{5.9}
    	\exp\left[ik_{0s}R\operatorname{sin}\left(\theta-\tau\right)\right].
    \end{equation}
    Therefore, the isolated saddle point is $\tau_0=\theta-\pi/2$.
    The contribution of $\tau_0$ in integral~%
    \eqref{eq:5.6} takes the form
    \begin{align}
    	\label{eq:5.10}
    	&H_{\varphi 0}^{(2)}\cong iqk_{0s}^{2}\operatorname{sin}\theta \nonumber \\
    	\tag{5.10}
    	&\left[\tilde\kappa M^{(i)}\xi_s\left(a,\theta\right)F_s\left(\theta\right)-\sum_{m=1}^{\infty}M_m\xi_m\left(a,\theta\right)F_m\left(\theta\right)\right]\frac{\exp\left(ik_{0s}R\right)}{R},
    \end{align}
    where
    \begin{equation}
    	\label{eq:5.11}
    	\tag{5.11}
    	\xi_s\left(a,\theta\right)=\frac{\tilde\chi_s}{k_{0s}\operatorname{sin}\theta}J_1\left(k_{0s}a\operatorname{sin}\theta\right)I_0\left(\tilde\chi_s a\right)-I_1\left(\tilde\chi_s a\right)J_0\left(k_{0s}a\operatorname{sin}\theta\right),
    \end{equation}
    \begin{equation}
	    \label{eq:5.12}
	    \tag{5.12}
	    \xi_m\left(a,\theta\right)=\frac{\chi_m}{k_{0s}\operatorname{sin}\theta}J_1\left(k_{0s}a\operatorname{sin}\theta\right)J_0\left(\chi_m a\right)-J_1\left(\chi_m a\right)J_0\left(k_{0s}a\operatorname{sin}\theta\right),
    \end{equation}
    \begin{equation}
    	\label{eq:5.13}
    	\tag{5.13}
    	F_s\left(\theta\right)=\frac{\beta^{-1}}{k_{0s}\left(\beta^{-2}-\operatorname{cos}^2\theta\right)},\;\;\;F_m\left(\theta\right)=\frac{k_{zm}}{k_{zm}^2-k_{0s}^2\operatorname{cos}^2\theta}.
    \end{equation}

    We note again that the electromagnetic field under consideration is the result of falling of TM$_{0s}$ mode at the open end of the waveguide.
    The frequency of this mode $\omega_{s}$ is determined by dispersion equation~%
    \eqref{eq:2.10}.
    Below, we present the results of numerical calculations of~%
    \eqref{eq:5.10}.
    The calculations were carried out for the incident mode corresponding the first solution of the dispersion equation, i.e. the mode with the lowest frequency.
    
\begin{figure*}[!h]
    	\includegraphics[width=13cm]{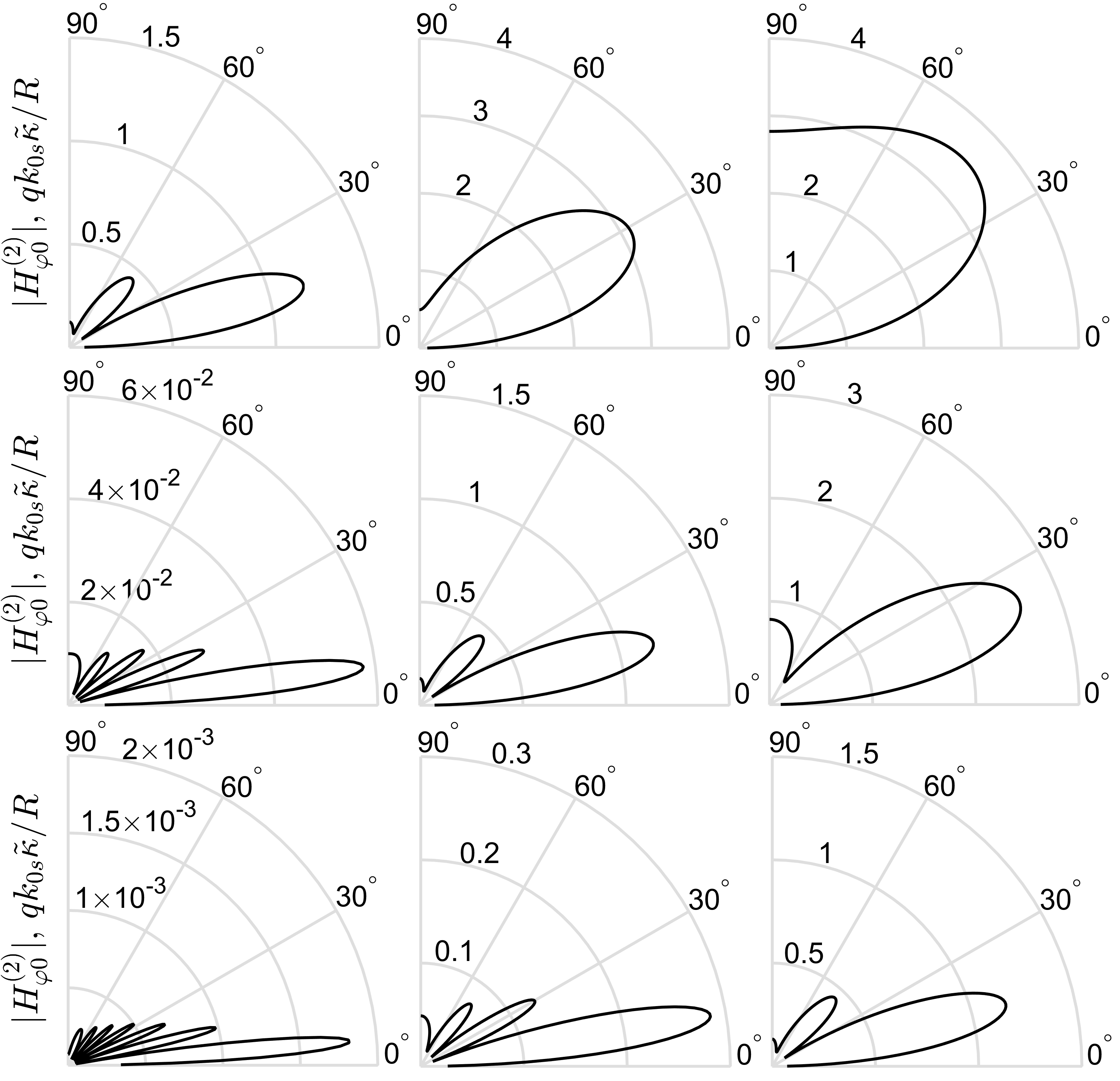}
    	\vspace{-10pt}
    	\caption{\label{Fig:2} The radiation patterns (in $qk_{0s}\tilde\kappa/R$ units) in the far-field zone of area $(2)$.
    		The bunch velocity is $\beta=0.9$.
    		The waveguide radius is $a=5$ cm (the top row), $a=10$ cm (the middle row) and $a=15$ cm (the bottom row).
    		The corrugation depth is $d_3=0.5$ cm (the left coloumn), $d_3=1$ cm (the middle coloumn) and $d_3=1.5$ cm (the right coloumn).
    		Other corrugation parameters: $d=0.1$ cm, $d_2=0.05$ cm.
    		The radiation frequency corresponds the first solution of dispersion equation~%
    		\eqref{eq:2.10}.
    	}
    \end{figure*}

\begin{figure*}[!h]
    	\includegraphics[width=13cm]{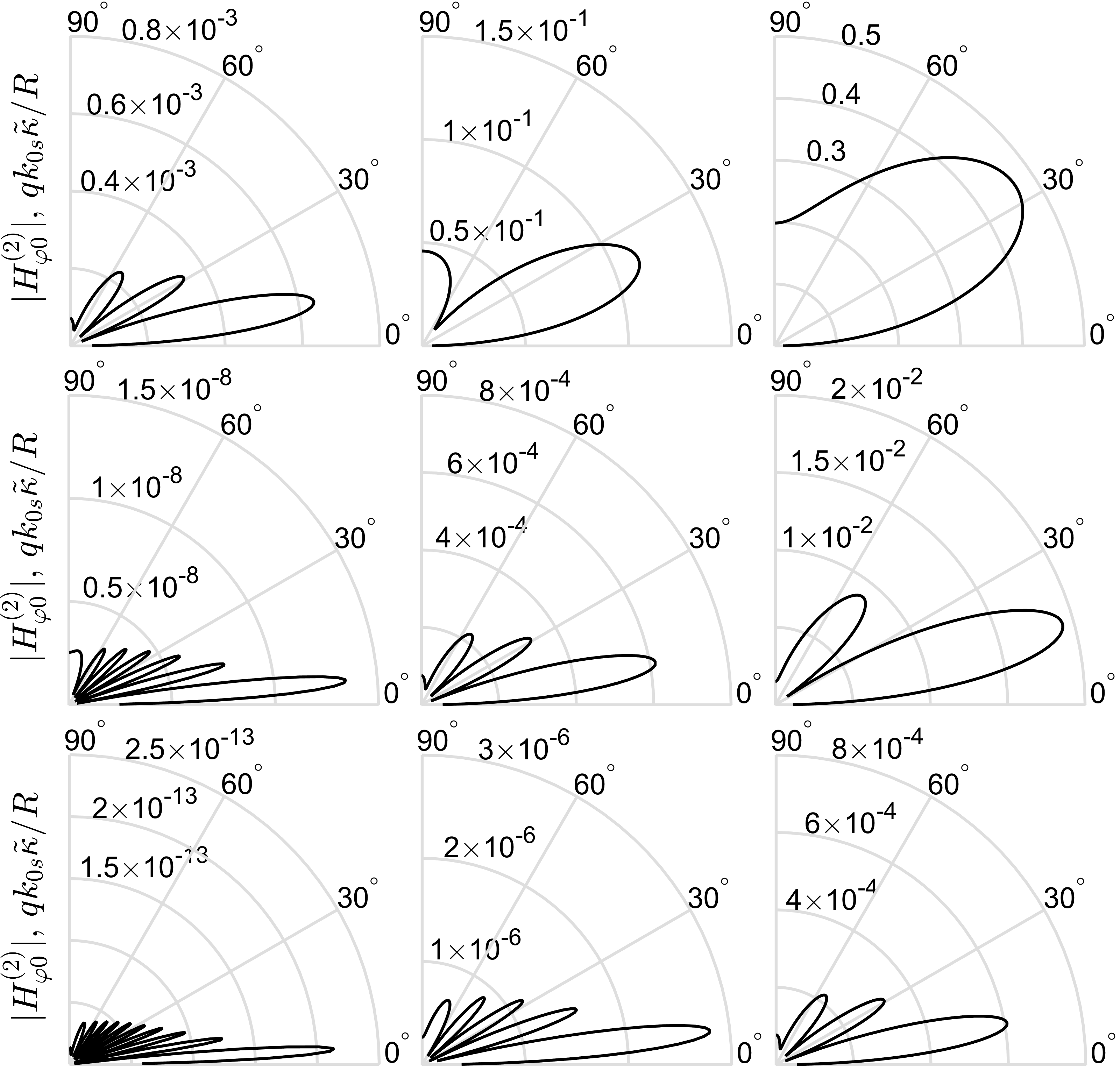}
    	\vspace{-10pt}
    	\caption{\label{Fig:3} The same as in Fig.~%
    		\ref{Fig:2} for $\beta=0.7$.
    	}
    \end{figure*}

    Figs.~%
    \ref{Fig:2} and~%
    \ref{Fig:3} illustrate the absolute value of~%
    \eqref{eq:5.10} depending on angle $\theta$ for the different values of bunch velocity $\beta$, corrugation depth $d_3$ and waveguide radius $a$.
    One can see that the field magnitude increases with increasing in the depth.
    This fact manifests most clearly for slower bunch and larger waveguide radius.
    As well, an increase in the field magnitude is observed with a decrease in the radius and an increase in the bunch velocity.
    As follows from the plots, the field distribution is characterized by the radiation maximums.
    The number of the maximums is proportional to the number of the propagating reflected modes~%
    \eqref{eq:2.11}, i.e. the ones having purely real longitudinal wavenumbers $k_{zm}$.
    The numerical calculations show that this number is greater for larger radius and smaller bunch velocity.
    The angle $\theta$ corresponding the direction of the radiation maximums decreases with decreasing in the corrugation depth and increasing in the radius.
    Under these conditions, the width of the maximums decreases as well, i.e. the lobes of the radiation patterns become narrower.
    For example, the direction of the main radiation maximum (with greater magnitude) for the problem parameters $\beta=0.9$, $a=15$ cm and $d_3=0.5$ cm is $\theta\cong 5^{\circ}$ and the angle increases to $\theta\cong 15^{\circ}$ with increasing in the depth to $d_3=1.5$ cm (the bottom row in Fig.~%
    \ref{Fig:2}).
    
\begin{figure*}[t]
    	\includegraphics[width=13cm]{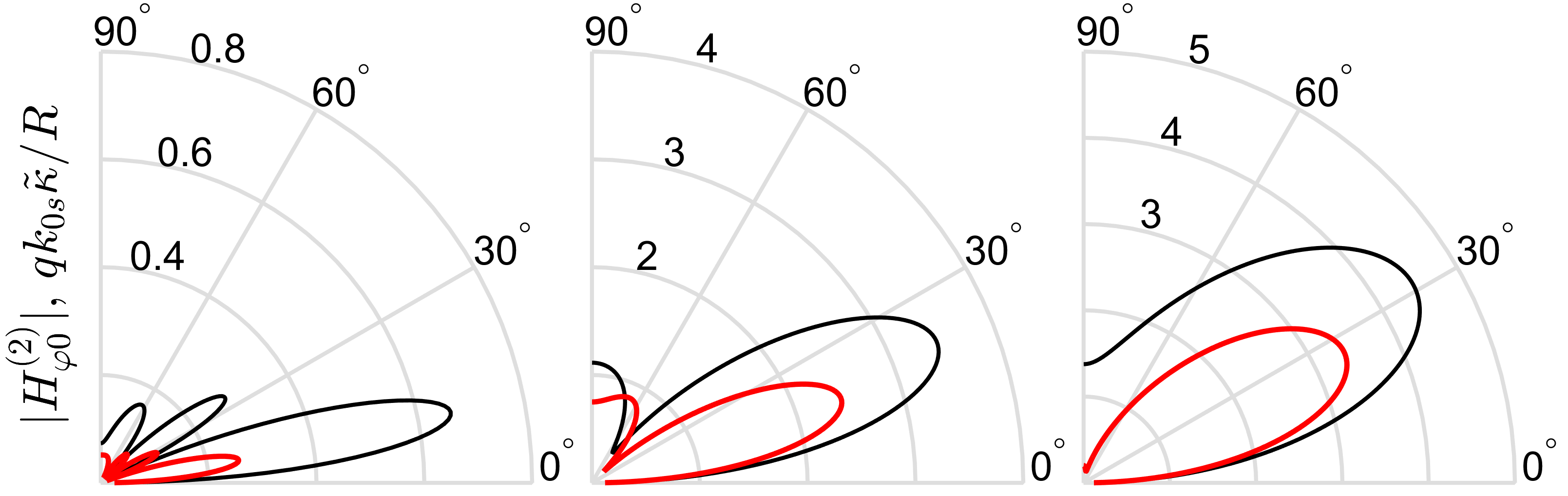}
    	\vspace{-10pt}
    	\caption{\label{Fig:4} The same as in the middle row of Fig.~%
    		\ref{Fig:2} for $d_2=0.07$ cm (the red curve) and $d_2=0.09$ cm (the black curve).
    	}
    \end{figure*}

    Let us also discuss the dependence of the field distribution on corrugation parameters $d$ and $d_2$, i.e. the corrugation period and the width of the corrugation grooves, respectively.
    The calculations show that, despite the relative smallness of these parameters in the ``longwave'' approximation, relation $d_2/d$ can strongly influence on the field distribution.
    Fig.~%
    \ref{Fig:4} illustrate the dependence of the radiation patterns on angle $\theta$ for $d_2/d=0.7$ (the red curve) and $d_2/d=0.9$ (the black curve).
    Other problem parameters are the same as in the bottom row of Fig.~%
    \ref{Fig:2}.
    As follows from Fig.~%
    \ref{Fig:4}, the field magnitude increases with an increase in $d_2/d$.
    This actually means that the radiation is more intensive when the corrugated structure is similar to the system of diaphragms ($d_2/d\rightarrow 1$).
    In particular, the analogous conclusion was made in work~%
    \cite{ESAV1} where we considered the shallow corrugated planar structure.
    \begin{figure*}[t]
    	\begin{center}
    		\includegraphics[]{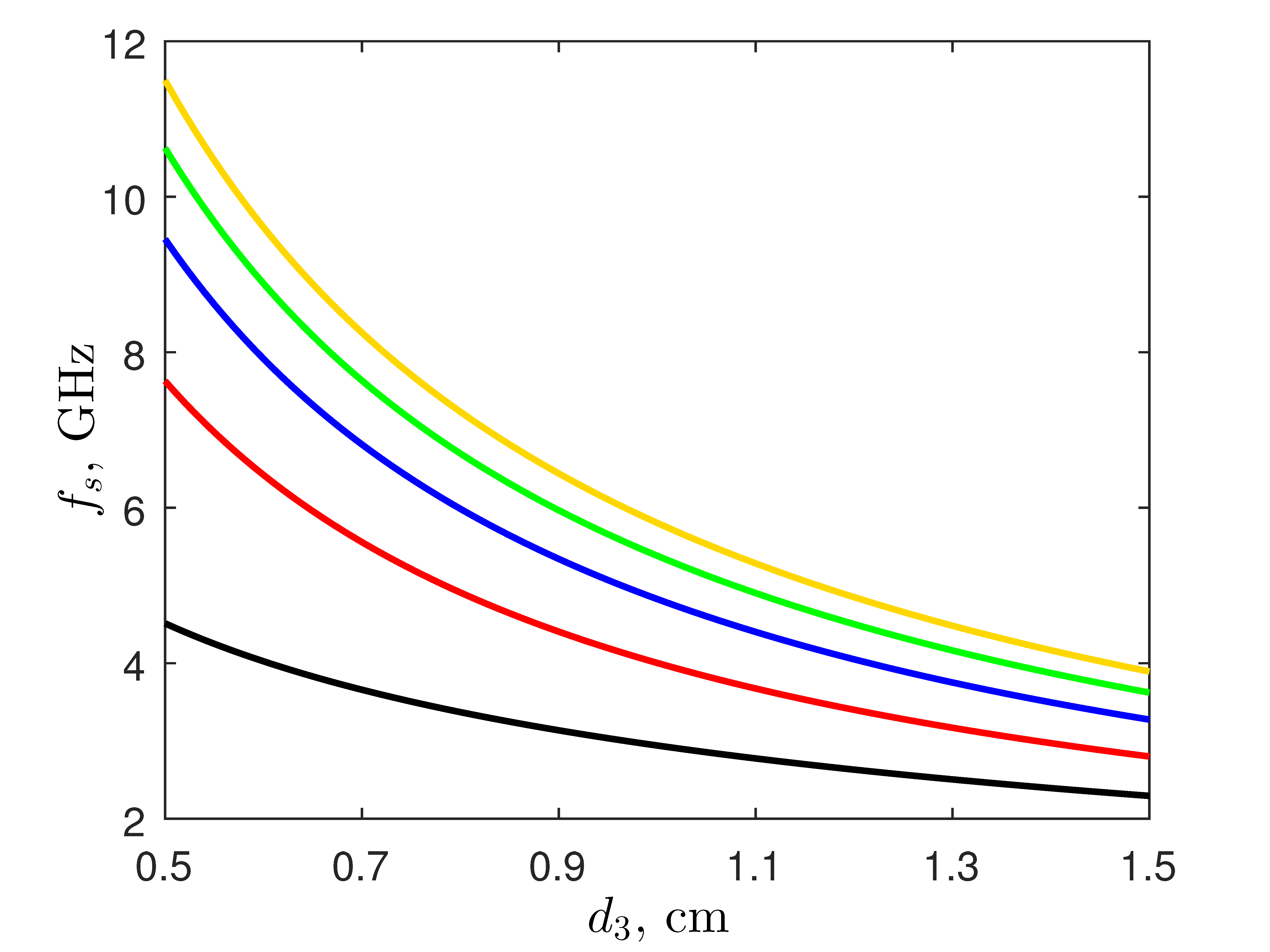}
    	\end{center}
    	\vspace{-15pt}
    	\caption{\label{Fig:5} The frequency $f_s$ (in GHz) of the incident mode depending on corrugation depth $d_3$.
    		The bunch velocity is $\beta=0.99$ (the black curve), $\beta=0.9$ (the red curve), $\beta=0.8$ (the blue curve), $\beta=0.7$ (the green curve) and $\beta=0.6$ (the yellow curve).
    		Other parameters: $a=5$ cm, $d=0.1$ cm, $d_2=0.05$ cm.
    	}
    \end{figure*}
    Finally, we present the dependence of the frequency $f_s=\omega_{s}/(2\pi)$, at which the incident TM$_{0s}$ mode is excited, on the corrugation depth for the different bunch velocities (see Fig.~%
    \ref{Fig:5}).
    As follows from the plots, the radiation is generated in the GHz range ($\approx 0.1$ THz).
    The frequency increases with decreasing in the depth and the velocity and, at the given parameters, varies by a factor of two or even three.
    Note that the dependence of the frequency on the waveguide radius is weak.
    This can be seen from dispersion equation~%
    \eqref{eq:2.10} which determines the radiation frequency.
    The solution of~%
    \eqref{eq:2.10} is mostly defined by impedance $\eta$ which does not depend on the radius.
    Thus, the frequency domain corresponding the results presented in Figs.~%
    \ref{Fig:2} and~%
    \ref{Fig:3} is approximately between the red and green curves in Fig.~%
    \ref{Fig:5}.

    \section{Conclusion}
    
    In this work, we have studied the diffraction process resulting from falling of the transverse magnetic mode at the open end of the cylindrical deeply corrugated waveguide.
    The corrugation period was assumed to be much less than the waveguide radius and the wavelengths under consideration (the ``longwave'' case).
    This assumption made it possible to apply the EBC method.
    In the approach presented, we have solved the corresponding Wiener-Hopf-Fock equation and constructed the infinite linear system for the reflection coefficients.
    This system was solved numerically using the reduction technique.
    The electromagnetic field outside the waveguide was investigated in the far-field zone.
    The corresponding radiation patterns were presented and analyzed.
    
    In particular, we have demonstrated that the radiation is more intensive for deeper corrugation and smaller waveguide size.
    Estimations show that it is possible to obtain the field magnitude comparable (or even greater) to the one in the case of the open-ended dielectric waveguide with a smooth wall~%
    \cite{IEEEopendw1,GalOpEndArx}.
    As follows from the radiation patterns, the corrugation similar to the system of diaphragms is the most effective for exciting the radiation.
    The frequency of this radiation increases significantly with decreasing in the corrugation depth and the bunch velocity, but practically does not depend on the wavegudie radius.
    The maximum value of the frequency is approximately $0.1$ THz at the parameters considered.
    This value can be increased in several times by varying the corrugation parameters.

\section{Acknowledgments}

This work was supported by the Russian Science Foundation (Grant No.~18-72-10137).

	\bibliographystyle{model1a-num-names}
	\bibliography{Manuscript}

\end{document}